\documentclass[reprint,
 amsmath,amssymb,
 aps,
]{revtex4-2}

\usepackage{graphicx}% Include figure files
\usepackage{dcolumn}% Align table columns on decimal point
\usepackage{bm}% bold math
\usepackage{xcolor}
\begin{document}

%\preprint{APS/123-QED}

\title{Spectral reorganization of space-time wave packets in presence of normal group-velocity dispersion}

\author{Layton A. Hall$^{1}$}
\author{Ayman F. Abouraddy$^{1,*}$}
%\email{raddy@creol.ucf.edu}
\affiliation{$^{1}$CREOL, The College of Optics \& Photonics, University of Central~Florida, Orlando, FL 32816, USA}
\affiliation{$^*$Corresponding author: raddy@creol.ucf.edu}

\begin{abstract}
Space-time wave packets (STWPs) are pulsed beams that propagate invariantly (without diffraction or dispersion) in linear media. The behavior of STWPs in free space is now well-established, and recently their propagation invariance was confirmed in both the normal and anomalous dispersion regimes. However, yet-to-be-observed rich dynamics of spectral reorganization have been predicted to occur in the presence of normal group-velocity dispersion (GVD). Indeed, propagation invariance in the normal-GVD regime is compatible with spatio-temporal spectra that are X-shaped, hyperbolic, parabolic, elliptical, or even separable along the spatial and temporal degrees-of-freedom. These broad varieties of field structures can be classified in a two-dimensional space parameterized by the group velocity of the STWP and its central axial wave number. Here we observe the entire span of spectral reorganization for STWPs in the paraxial regime in normally dispersive ZnSe at a wavelength $\sim1$~$\mu$m with STWPs of on-axis pulse width of $\sim\!200$~fs. By tuning the group velocity and central axial wave number of the STWP, we observe transitions in the structure of the spatio-temporal spectrum and verify the associated change in its intensity profile. These results lay the foundation for initiating novel phase-matched nonlinear processes using STWPs in dispersive media. 
\end{abstract}

%\setboolean{displaycopyright}{true}

\maketitle

\section{Introduction}

Space-time wave packets (STWPs) are a class of spatio-temporally structured pulsed optical fields that are propagation-invariant in linear media \cite{Yessenov22AOP,Kondakci16OE,Parker16OE,Wong17ACSP2,Porras17OL,Efremidis17OL}. To date, STWPs have been experimentally studied predominantly in free space \cite{Kondakci17NP} and non-dispersive dielectrics \cite{Bhaduri19Optica}, in which a host of useful characteristics of STWPs have been verified, including: tunable group velocities \cite{Salo01JOA,Wong17ACSP2,Kondakci19NC} and acceleration \cite{Clerici08OE,ValtnaLukner09OE,Yessenov20PRL2,Li20SR,Li20CP,Li21CP,Hall22OLArbAccel}, self-healing \cite{Kondakci18OL}, and anomalous refraction \cite{Bhaduri20NP,Motz21OL}. Moreover, almost arbitrary dispersion profiles can be introduced in free space \cite{Yessenov21ACSP,Hall21OL3NormalGVD}, which has enabled the observation of the space-time Talbot effect \cite{Hall21APLP}.

Although the propagation invariance of STWPs in linear dispersive media has been pursued theoretically \cite{Porras01OL,Porras03OL,Porras03PRE2,Longhi04OL,Porras04PRE}, the only experimental test has been the propagation of X-waves in normally dispersive silica glass \cite{Sonajalg96OL,Sonajalg97OL} (subsequent attempts \cite{Dallaire09OE,Jedrkiewicz13OE} with STWPs could not confirm this effect). Extensive studies have investigated the \textit{generation} of STWPs in nonlinear dispersive media \cite{DiTrapani03PRL,Faccio07OE}, but their propagation invariance was not verified. However, rapidly evolving developments in preparing STWPs via universal angular-dispersion synthesis \cite{Hall21OEUniversal} has very recently made it possible to confirm dispersion-free propagation of STWPs in dispersive media -- in presence of either normal or anomalous GVD \cite{Hall22LPRdispersion}.

In addition to propagation invariance, a rich spatio-temporal structural dynamics has been predicted to accompany STWPs in dispersive media, which have \textit{not} been observed to date. In free space, although elliptic, hyperbolic, or parabolic spatio-temporal spectra can occur as the STWP group velocity is tuned, nevertheless, they are all well-approximated in the paraxial regime by a parabola, and their intensity profiles usually show no accompanying variation (except in presence of spectral-phase modulation \cite{Kondakci18PRL,Wong21OE}). In contrast, Malaguti and Trillo (MT henceforth) have predicted fascinating changes in the spatio-temporal spectra of STWPs in the presence of normal GVD \cite{Malaguti09PRA} (see also \cite{Lotti10PRA}, in addition to earlier \cite{Porras01OL,Porras03OL,Porras03PRE2,Longhi04OL,Porras04PRE} and recent work \cite{Bejot22ACSP}). Here, a wide range of spatio-temporal spectra are compatible with dispersion-free propagation in the normal-GVD regime: including elliptic, parabolic, hyperbolic, X-shaped, and separable spectra -- all clearly observable in the paraxial regime. Tuning the STWP parameters lead to crossovers between these different structures and thus `spectral reorganization' of the STWP. Once the medium characteristics are fixed, two key parameters dictate these dynamics: the STWP group velocity and its central axial wave number. Tuning one of these parameters or both leads to dramatic changes in the spectral structure of the STWP if it is to remain dispersion-free in the medium. Indeed, tuning these parameters is predicted to drive structural reorganization in the STWP spectrum and spatio-temporal intensity profile \cite{Malaguti09PRA}.

Here, we observe for the first time the broad range of predicted spectral reorganization of propagation-invariant STWPs taking place in presence of normal GVD as the STWP group velocity and central axial wave number are continuously tuned within the paraxial regime. Starting with generic femtosecond pulses at a wavelength $\lambda_{\mathrm{o}}\!\approx\!1$~$\mu$m, we synthesize STWPs with on-axis pulsewidths in the range $\sim\!200-450$~fs (bandwidths of $8-16$~nm), which are propagation invariant in normally dispersive ZnSe. By tuning the group velocity and central axial wave number, we span the two-dimensional (2D) parameter space proposed by MT (which we refer to as the MT-plane), and reveal the wealth of spatio-temporal spectral structures and profiles for STWPs that are compatible with propagation invariance in ZnSe; including elliptic, parabolic, hyperbolic (with gaps along the spatial or temporal spectral axes), X-shaped, and even approximately separable spatio-temporally spectra. In light of the importance of dispersion in the phase-matching of a wide variety of nonlinear optical processes, we expect that these results will be useful in producing new nonlinear and quantum optical effects in bulk media and in planar nonlinear cavities and waveguides \cite{Shiri20NC,Kibler2021PRL,Guo21PRR,Bejot2021ACSP,Bejot22ACSP}.

\section{Theory of Propagation-invariant space-time wave packets in presence of normal GVD}

\subsection{Spectral representation of conventional pulsed beams on the light-cone surface}

In a medium of refractive index $n(\omega)$, the dispersion relationship for a monochromatic plane wave is $k_{x}^{2}+k_{z}^{2}\!=\!k^{2}$, where $k(\omega)\!=\!n(\omega)\tfrac{\omega}{c}$ is the wave number, $\omega$ is the angular temporal frequency, $c$ is the speed of light in vacuum, and $k_{x}$ and $k_{z}$ are the transverse and longitudinal components of the wave vector along the transverse $x$ and axial $z$ coordinates, respectively (we hold the field uniform along $y$ for simplicity). We expand $k(\omega)\!=\!k(\omega_{\mathrm{o}}+\Omega)$ to second order in $\Omega\!=\!\omega-\omega_{\mathrm{o}}$ (where $\omega_{\mathrm{o}}$ is a fixed frequency): $k(\omega)\!=\!n_{\mathrm{m}}k_{\mathrm{o}}+\tfrac{\Omega}{\widetilde{v}_{\mathrm{m}}}+\tfrac{1}{2}k_{2}\Omega^{2}$, where $n_{\mathrm{m}}\!=\!n(\omega_{\mathrm{o}})$, $k_{\mathrm{o}}\!=\!\tfrac{\omega_{\mathrm{o}}}{c}$, $\widetilde{v}_{\mathrm{m}}\!=\!1\big/\tfrac{dk}{d\omega}\big|_{\omega_{\mathrm{o}}}$ is the group velocity of a plane-wave pulse in the medium, and $k_{2}\!=\!\tfrac{d^{2}k}{d\omega^{2}}\big|_{\omega_{\mathrm{o}}}$ is the GVD coefficient (positive-valued for normal GVD) \cite{SalehBook07}. Writing the field as $E(x,z;t)\!=\!e^{i(n_{\mathrm{m}}k_{\mathrm{o}}z-\omega_{\mathrm{o}}t)}\psi(x,z;t)$, the envelope is:
\begin{equation}\label{Eq:GeneralPulsedBeam}
\psi(x,z;t)\!=\!\!\iint\!dk_{x}d\Omega\widetilde{\psi}(k_{x},\Omega)e^{i(k_{x}x-\tfrac{k_{x}^{2}z}{2n_{\mathrm{m}}k_{\mathrm{o}}})}e^{-i\Omega(t-\tfrac{z}{\widetilde{v}_{\mathrm{m}}})}e^{ik_{2}\Omega^{2}z/2},
\end{equation}
where the spatio-temporal spectrum $\widetilde{\psi}(k_{x},\Omega)$ is the Fourier transform of $\psi(x,0;t)$, and the envelope undergoes diffractive spatial spreading and dispersive temporal broadening with propagation along $z$. With respect to the geometric cone $k_{x}^{2}+k_{z}^{2}\!=\!(\tfrac{\omega}{c})^{2}$ associated with propagation in free space [Fig.~\ref{Fig:LightCones}(a)], the light-cone $k_{x}^{2}+k_{z}^{2}\!=\!(n(\omega)\tfrac{\omega}{c})^{2}$ in presence of normal GVD is curved inwardly [Fig.~\ref{Fig:LightCones}(c)], a fact that undergirds many of the phenomena described below. For a conventional pulsed beam (Eq.~\ref{Eq:GeneralPulsedBeam}), the spectral support for $\widetilde{\psi}(k_{x},\Omega)$ on the light-cone surface is a 2D domain, corresponding to finite spatial and temporal bandwidths $\Delta k_{x}$ and $\Delta\omega$, respectively.

\subsection{Spectral representation of STWPs on the light-cone surface}

\begin{figure}[t!]
\centering
\includegraphics[width=8.6cm]{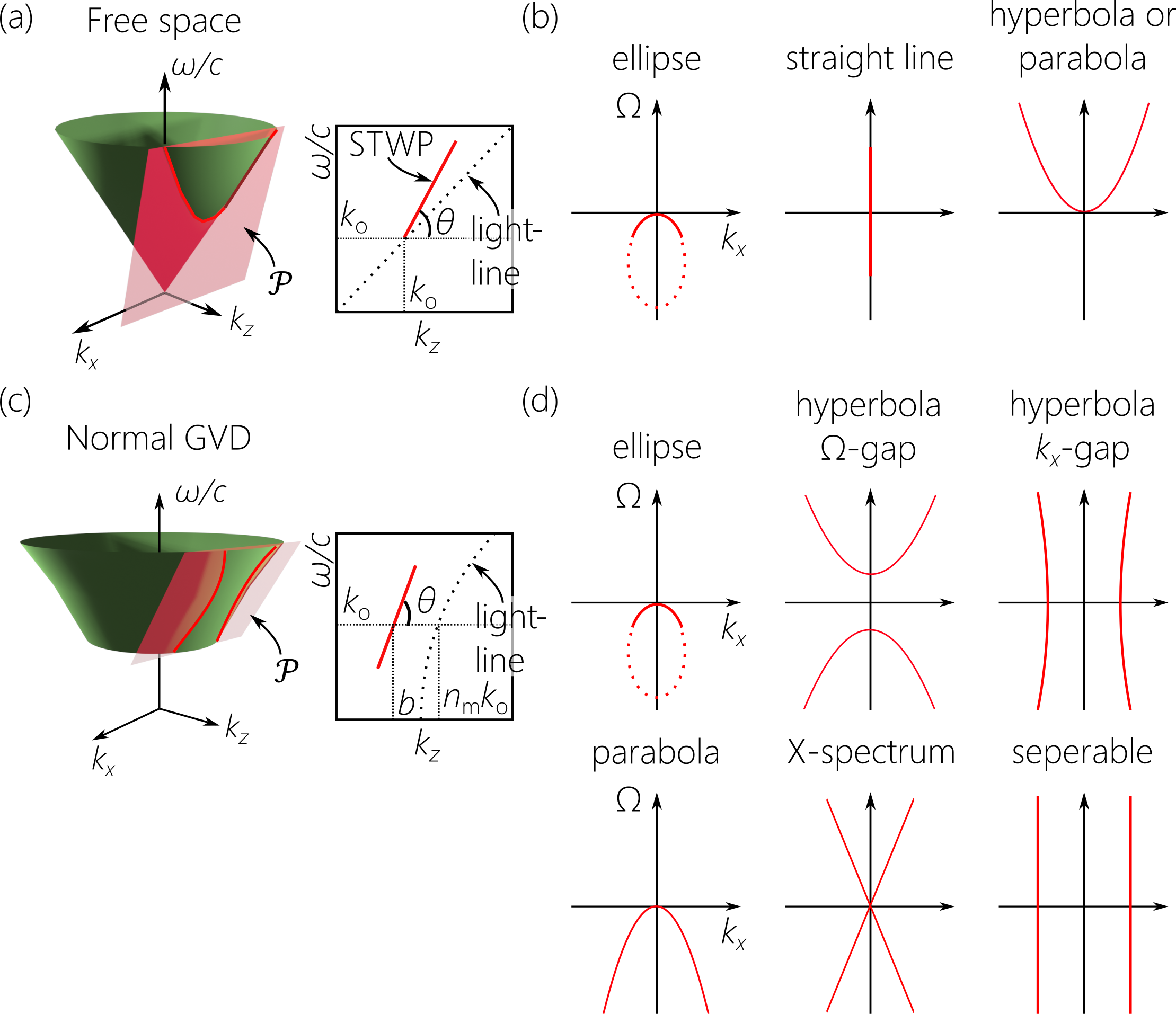}
\caption{(a) The spectral support of a free-space STWP is the intersection of the light-cone $k_{x}^{2}+k_{z}^{2}\!=\!(\tfrac{\omega}{c})^{2}$ with a tilted spectral plane $\mathcal{P}$. (b) Possible spectral projections onto the $(k_{x},\Omega)$-plane in free space. (c) The light-cone $k_{x}^{2}+k_{z}^{2}\!=\!(n(\omega)\tfrac{\omega}{c})^{2}$ in presence of normal GVD and the spectral support of a STWP at its intersection with $\mathcal{P}$. (d) Possible spectral projections onto the $(k_{x},\Omega)$-plane in presence of normal GVD.}
\label{Fig:LightCones}
\end{figure}

The impact of both diffraction and dispersion can be halted in a STWP by imposing a strict association between $k_{x}$ and $\Omega$, so that the spectral support on the light-cone surface becomes a 1D curve rather than a 2D domain \cite{Donnelly93ProcRSLA,FigueroaBook14,Yessenov19PRA,Yessenov22AOP}: $\widetilde{\psi}(k_{x},\Omega)\!\rightarrow\!\widetilde{\psi}(\Omega)\delta(k_{x}-k_{x}(\Omega))$. The association $k_{x}\!=\!k_{x}(\Omega)$ enforces the following constraint on the axial wave number \cite{FigueroaBook14}:
\begin{equation}\label{Eq:GeneralConstraintKz}
k_{z}(\Omega)=b+\frac{\Omega}{\widetilde{v}},
\end{equation}
where $\widetilde{v}$ is the group velocity of the STWP in the dispersive medium ($\widetilde{v}$ may be higher or lower than $\widetilde{v}_{\mathrm{m}}$), and $b\!<\!n_{\mathrm{m}}k_{\mathrm{o}}$ is the axial wave number at $\omega\!=\!\omega_{\mathrm{o}}$ ($\Omega\!=\!0$), referred to henceforth as the central axial wave number. Writing $b\!=\!n_{\mathrm{m}}k_{\mathrm{o}}\cos{\varphi_{\mathrm{o}}}$, $\varphi_{\mathrm{o}}$ is the angle the frequency component $\omega_{\mathrm{o}}$ makes with the $z$-axis, and $b\!<\!n_{\mathrm{m}}k_{\mathrm{o}}$ therefore indicates an angular or directional offset at $\omega_{\mathrm{o}}$. The constraint on $k_{z}$ in Eq.~\ref{Eq:GeneralConstraintKz} confines the STWP spectral support on the dispersive light-cone surface to its intersection with a spectral plane $\mathcal{P}$ defined by Eq.~\ref{Eq:GeneralConstraintKz}. This plane is parallel to the $k_{x}$-axis and makes an angle $\theta$ with respect to the $k_{z}$-axis, where $\widetilde{v}\!=\!c\tan{\theta}$. This intersection is in general a conic section (if terms up to second order in $\Omega$ are retained), and its projection onto the $(k_{z},\tfrac{\omega}{c})$-plane is a straight line, indicating absence of dispersion of any order for the STWP in the dispersive medium [Fig.~\ref{Fig:LightCones}(c)].

In presence of normal GVD, a wide variety of geometric structures can occur for the spatio-temporal spectra (or `spectra' for brevity), including: an ellipse; a hyperbola (both branches of which can be accessible in the paraxial regime) with a gap along $k_{x}$ or along $\Omega$; a parabola; an X-shaped spectrum; or a spectrum that is separable in terms of $k_{x}$ and $\Omega$ formed of a pair of parallel straight lines [Fig.~\ref{Fig:LightCones}(d)]. In contrast, the spectra in free space can only be an ellipse, a hyperbola (only a single branch of which is accessible for $\omega\!>\!0$), a parabola, or a straight line ($\mathcal{P}$ is tangential to the light-cone); see Fig.~\ref{Fig:LightCones}(b). In the paraxial regime, the portion of the spectrum in free space in the vicinity of $k_{x}\!=\!0$ (so-called `baseband' STWPs \cite{Yessenov19PRA}) can always be approximated by a parabola \cite{Yessenov22AOP}.

\subsection{Representation in the MT-plane}

\begin{figure*}[t!]
\centering
\includegraphics[width=17.6cm]{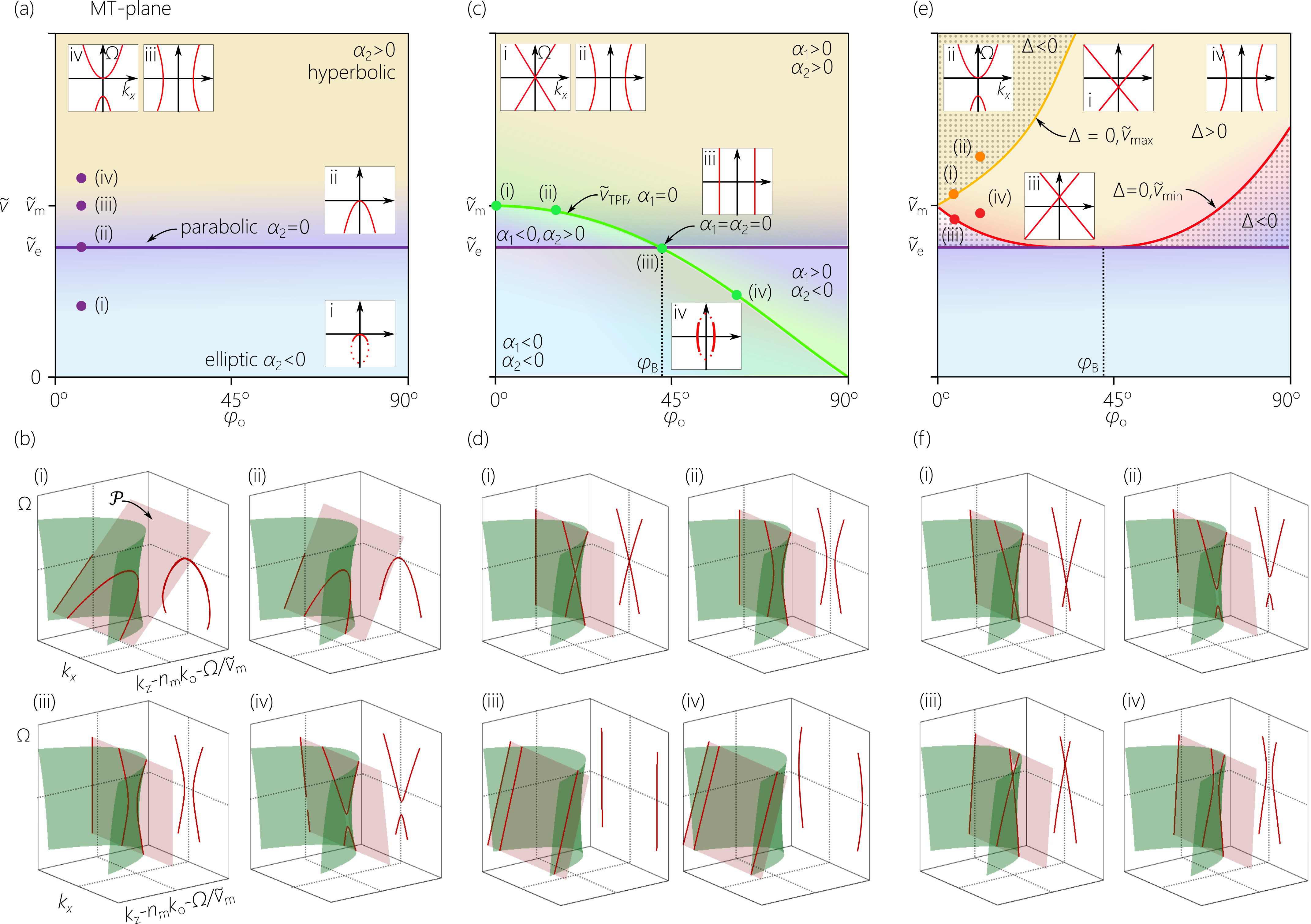}
\caption{(a) In the MT-plane spanned by $\varphi_{\mathrm{o}}$ (the angular offset at $\omega\!=\!\omega_{\mathrm{o}}$) and the STWP group velocity $\widetilde{v}$, each point determines a unique trio of coefficients $\alpha_{0}$, $\alpha_{1}$, and $\alpha_{2}$ (Eq.~\ref{Eq:kx_general}). The horizontal line $\widetilde{v}\!=\!\widetilde{v}_{\mathrm{e}}\!<\!\widetilde{v}_{\mathrm{m}}$ is the locus of $\alpha_{2}\!=\!0$ (parabolic spectra), separating the plane into an upper region $\alpha_{2}\!>\!0$ (hyperbolic spectra), and a lower region $\alpha_{2}\!<\!0$ (elliptical spectra). (b) Representation on the dispersive light-cone of the STWPs identified by points (i)-(iv) in (a). (c) The curve $\widetilde{v}\!=\!\widetilde{v}_{\mathrm{TPF}}(\varphi_{\mathrm{o}})$ is the locus of $\alpha_{1}\!=\!0$. The point $\varphi_{\mathrm{o}}\!=\!\varphi_{\mathrm{B}}$ ($\alpha_{1}\!=\!\alpha_{2}\!=\!0$) at $\widetilde{v}\!=\!\widetilde{v}_{\mathrm{e}}\!=\!\widetilde{v}_{\mathrm{TPF}}(\varphi_{\mathrm{B}})$ corresponds to a dispersion-free pulsed Bessel beam that is separable with respect to space and time. (d) Spectral representation of the STWPs identified by points (i)-(iv) in (c). (e) The domain $\alpha_{2}\!>\!0$ of hyperbolic spectra is subdivided by means of the curves $\widetilde{v}_{\mathrm{max}}$ and $\widetilde{v}_{\mathrm{min}}$ corresponding to $\Delta\!=\!0$, along which the spectrum is X-shaped. In the shaded regions where $\Delta\!>\!0$, the hyperbolas have an $\Omega$-gap, and a $k_{x}$-gap in the region between $\widetilde{v}_{\mathrm{min}}$ and $\widetilde{v}_{\mathrm{max}}$ where $\Delta\!<\!0$. (f) Spectral representation of the STWPs identified by points (i)-(iv) in (e).}
\label{Fig:MTPlaneStructure}
\end{figure*}

The field in the dispersive medium after applying the constraint in Eq.~\ref{Eq:GeneralConstraintKz} is $E(x,z;t)\!=\!e^{i(bz-\omega_{\mathrm{o}})}\psi(x,z;t)$, where:
\begin{equation}
\psi(x,z;t)=\!\int\!\!d\Omega\;\widetilde{\psi}(\Omega)e^{ik_{x}(\Omega)x}e^{-i\Omega(t-\tfrac{z}{\widetilde{v}})}=\psi(x,0;t-z/\widetilde{v}),
\end{equation}
which represents a STWP traveling rigidly (without dispersion or diffraction) at a group velocity $\widetilde{v}$. The dispersion relationship for propagation invariance in the medium is $k_{x}^{2}+(b+\tfrac{\Omega}{\widetilde{v}})^{2}\!=\!(n_{\mathrm{m}}k_{\mathrm{o}}+\tfrac{\Omega}{\widetilde{v}_{\mathrm{m}}}+\tfrac{1}{2}k_{2}\Omega^{2})^{2}$, and retaining terms up to second order in $\Omega$ yields:
\begin{equation}\label{Eq:kx_general}
k_{x}^{2}(\Omega)=\alpha_{0}+2\alpha_{1}\Omega+\alpha_{2}\Omega^{2},
\end{equation}
where the coefficients $\alpha_{0}$, $\alpha_{1}$, and $\alpha_{2}$ are given by:
\begin{eqnarray}
\alpha_{0}&=&(n_{\mathrm{m}}k_{\mathrm{o}}\sin{\varphi_{\mathrm{o}}})^{2},\nonumber\\
\alpha_{1}&=&n_{\mathrm{m}}k_{\mathrm{o}}\cos{\varphi_{o}}\left(\frac{1}{\widetilde{v}_{\mathrm{TPF}}(\varphi_{\mathrm{o}})}-\frac{1}{\widetilde{v}}\right),\nonumber\\
\alpha_{2}&=&\frac{1}{\widetilde{v}_{\mathrm{e}}^{2}}-\frac{1}{\widetilde{v}^{2}};
\end{eqnarray}
where $\widetilde{v}_{\mathrm{TPF}}(\varphi_{\mathrm{o}})\!=\!\widetilde{v}_{\mathrm{m}}\cos{\varphi_{\mathrm{o}}}$, $\widetilde{v}_{\mathrm{e}}\!=\!\widetilde{v}_{\mathrm{m}}\cos{\varphi_{\mathrm{B}}}$, and $\tan^{2}{\varphi_{\mathrm{B}}}\!=\!n_{\mathrm{m}}k_{\mathrm{o}}k_{2}\widetilde{v}_{\mathrm{m}}^{2}$ (the physical significance of these terms will become clear shortly). It is crucial to appreciate that Eq.~\ref{Eq:kx_general} is \textit{not} a perturbative expansion of $k_{x}$ in terms of $\Omega$ (although we do truncate the equation at second order in $\Omega$). Indeed, \textit{any} of the terms in Eq.~\ref{Eq:kx_general} can vanish or dominate the sum. In fact, $k_{x}$ may not even be differentiable with respect to $\Omega$ at $\Omega\!=\!0$.

Once the material properties $n_{\mathrm{m}}$, $\widetilde{v}_{\mathrm{m}}$, and $k_{2}$ are determined at $\omega_{\mathrm{o}}$, the structure of the dispersive light-cone is fixed. Only two independent parameters -- $\widetilde{v}$ and $\varphi_{\mathrm{o}}$ -- remain to determine the values of the coefficients $\alpha_{0}$, $\alpha_{1}$, and $\alpha_{2}$. Therefore, all classes of propagation-invariant STWPs in the normal-GVD regime can be uniquely mapped to a 2D parameter space spanned by $\widetilde{v}$ and $\varphi_{\mathrm{o}}$, which we refer to as the MT-plane \cite{Malaguti09PRA}. Each point in this space identifies a class of STWPs associated with a particular functional form of $k_{x}(\Omega)$ in Eq.~\ref{Eq:kx_general}. The horizontal axis at $\widetilde{v}\!=\!0$ represents monochromatic beams, whereas the vertical axis at $\varphi_{\mathrm{o}}\!=\!0$ ($b\!=\!n_{\mathrm{m}}k_{\mathrm{o}}$) corresponds to STWPs whose spectra pass through the point $(k_{x},\Omega)\!=\!(0,0)$. 

In contrast, free-space STWPs are uniquely identified by the single parameter $\widetilde{v}$, and hence the fewer spectral structures accessible in that case \cite{Kondakci17NP} [compare Fig.~\ref{Fig:LightCones}(b) with Fig.~\ref{Fig:LightCones}(d)]. We proceed to describe the changes in the spectral structure for STWPs across the MT-plane.

\subsection{Escape velocity}

A critical class of STWPs are those whose group velocity is equal to the `escape' velocity $\widetilde{v}_{\mathrm{e}}\!=\!\widetilde{v}_{\mathrm{m}}\cos{\varphi_{\mathrm{B}}}\!<\!\widetilde{v}_{\mathrm{m}}$, whereupon the coefficient $\alpha_{2}$ vanishes. The locus of STWPs in the MT-plane for which $\alpha_{2}\!=\!0$ is thus the horizontal line $\widetilde{v}\!=\!\widetilde{v}_{\mathrm{e}}$ [Fig.~\ref{Fig:MTPlaneStructure}(a)], which divides the MT-plane into two regions: the spectrum is a hyperbola in the upper region ($\widetilde{v}\!>\!\widetilde{v}_{\mathrm{e}}$ and $\alpha_{2}\!>\!0$), and is an ellipse in the lower region ($\widetilde{v}\!<\!\widetilde{v}_{\mathrm{e}}$ and $\alpha_{2}\!<\!0$). Along $\widetilde{v}\!=\!\widetilde{v}_{\mathrm{e}}$ where $\alpha_{2}\!=\!0$, the spectrum $k_{x}^{2}(\Omega)\!=\!\alpha_{\mathrm{o}}+2\alpha_{1}\Omega$ is a parabola [Fig.~\ref{Fig:MTPlaneStructure}(b)]. Leaving the paraxial regime, the peak of the parabola reaches $\Omega\!\rightarrow\!\infty$ when $\varphi_{\mathrm{o}}\!=\!\varphi_{\mathrm{B}}$, whereupon the spectrum becomes a pair of vertical lines. Subsequently, when $\varphi_{\mathrm{o}}\!>\!\varphi_{\mathrm{B}}$, the orientation of the parabola flips, and its peak lies at $\Omega\!>\!0$. However, these dynamics around $\varphi_{\mathrm{o}}\!=\!\varphi_{\mathrm{B}}$ occur typically in the non-paraxial regime.

\subsection{Tilted-pulse-front velocity}

A second constraint of interest is that resulting in $\widetilde{v}\!=\!\widetilde{v}_{\mathrm{TPF}}(\varphi_{\mathrm{o}})\!=\!\widetilde{v}_{\mathrm{m}}\cos{\varphi_{\mathrm{o}}}$, where $\alpha_{1}\!=\!0$ and is represented by the curved path in Fig.~\ref{Fig:MTPlaneStructure}(c). The form of this group velocity $\widetilde{v}_{\mathrm{TPF}}(\varphi_{\mathrm{o}})$ is that associated with so-called tilted pulse fronts (TPFs) in the dispersive medium \cite{Hebling96OQE,Szatmari96OL}. A TPF results from introducing conventional angular dispersion into a pulsed field (as occurs when it traverses a grating or a prism \cite{Fulop10Review,Torres10AOP}).

A wide variety of spectra arise along this curve. At $\varphi_{\mathrm{o}}\!=\!0$ and $\widetilde{v}_{\mathrm{TPF}}(0^{\circ})\!=\!\widetilde{v}_{\mathrm{m}}\!>\!\widetilde{v}_{\mathrm{e}}$ ($\alpha_{2}\!>\!0$), we have $\alpha_{0}\!=\!\alpha_{1}\!=\!0$ and thus $k_{x}\!=\!\pm\sqrt{\alpha_{2}}\Omega$, so that the spectrum is X-shaped and formed of two straight lines symmetrically tilted and intersecting at $(k_{x},\Omega)\!=\!(0,0)$, which is associated with $\mathcal{P}$ tangential to the light-cone at $\omega_{\mathrm{o}}$ [Fig.~\ref{Fig:MTPlaneStructure}(d)]. The special case of an X-shaped spectrum was studied for free-space STWPs where it was demonstrated that they experience anomalous-GVD \cite{Hall21PRA} (and thus may propagate dispersion-free in the normal-GVD regime).

As $\varphi_{\mathrm{o}}$ increases in the $\alpha_{2}\!>\!0$ region ($\widetilde{v}\!>\!\widetilde{v}_{\mathrm{e}}$), the X-shaped spectrum evolves into a hyperbola with a gap along $k_{x}$, and whose center is at $(k_{x},\Omega)\!=\!(0,0)$ [Fig.~\ref{Fig:MTPlaneStructure}(d)]. Upon reaching the point $\widetilde{v}\!=\!\widetilde{v}_{\mathrm{e}}$ at $\varphi_{\mathrm{o}}\!=\!\varphi_{\mathrm{B}}$ whereupon $\alpha_{1}\!=\!\alpha_{2}\!=\!0$ and $k_{x}\!=\!\pm n_{\mathrm{m}}k_{\mathrm{o}}\sin{\varphi_{\mathrm{B}}}$, the hyperbolic curvature vanishes and the spectrum is formed of two parallel straight lines, thus decoupling $k_{x}$ and $\Omega$ [Fig.~\ref{Fig:MTPlaneStructure}(d)]. The envelope $\psi_{\mathrm{B}}(x,z;t)\!\propto\!\cos{(k_{x}x)}\psi_{t}(t-z/\widetilde{v}_{\mathrm{e}})$ becomes \textit{separable} in space and time, corresponding to a cosine wave in the transverse plane and an arbitrary pulse profile in time propagating dispersion-free at a group velocity $\widetilde{v}_{\mathrm{e}}$. When both transverse dimensions are taken into consideration, the transverse profile takes the form of a \textit{Bessel} beam $\psi_{\mathrm{B}}(r,z;t)\!\propto\!J_{0}(k_{r}r)\psi_{t}(t-z/\widetilde{v}_{\mathrm{e}})$, where $r$ is the radial coordinate and $k_{r}$ a fixed radial wave number. This STWP in presence of normal GVD has been studied theoretically \cite{Campbell90JASA,Liu98JMO,Hu02JOSAA,Lu03JOSAA} and is typically called a pulsed Bessel beam \cite{Turunen10PO}. To the best of our knowledge, it has not been realized experimentally to date because $\varphi_{\mathrm{B}}$ is typically in the non-paraxial domain (e.g., $\varphi_{\mathrm{B}}\!\approx\!19.2^{\circ}$ for ZnSe at $\lambda_{\mathrm{o}}\!\approx\!1$~$\mu$m). However, prior to reaching $\varphi_{\mathrm{B}}$, the hyperbolic curvature is low and the STWP is approximately separable (see below). When $\widetilde{v}_{\mathrm{TPF}}\!<\!\widetilde{v}_{\mathrm{e}}$ and $\alpha_{2}\!<\!0$ ($\varphi_{\mathrm{o}}\!>\!\varphi_{\mathrm{B}}$), the spectrum is an ellipse whose center is $(k_{x},\Omega)\!=\!(0,0)$; the ellipse is deep within the non-paraxial regime in the vicinity of $\omega_{\mathrm{o}}$ [Fig.~\ref{Fig:MTPlaneStructure}(d)].

\subsection{Increasing the group velocity at zero angular offset}

The physical meaning of the escape velocity $\widetilde{v}_{\mathrm{e}}$ can be understood geometrically by following the evolution of the spectrum as we increase $\widetilde{v}$ while holding $\varphi_{\mathrm{o}}\!=\!0^{\circ}$ ($\alpha_{0}\!=\!0$). At $\widetilde{v}\!=\!0$, $\mathcal{P}$ is a horizontal iso-frequency plane that intersects with the light-cone in a circle (a monochromatic beam). As $\widetilde{v}$ increases, the intersection becomes an ellipse with vertex at $(k_{x},\Omega)\!=\!(0,0)$, while the opposing vertex moves downwards along the $\Omega$-axis with further increase in $\widetilde{v}$ until $\widetilde{v}\!=\!\widetilde{v}_{\mathrm{e}}$ ($\alpha_{2}\!=\!0$), at which point $\mathcal{P}$ `escapes' from re-intersecting with the light-cone, and the spectrum becomes a parabola; hence the appellation `escape velocity'.

In the region $\widetilde{v}_{\mathrm{e}}\!<\!\widetilde{v}\!<\!\widetilde{v}_{\mathrm{m}}$ ($\alpha_{2}\!>\!0$ but $\alpha_{1}\!<\!0$), the spectrum is a hyperbola with a gap along $\Omega$; the vertex of the lower branch is $(k_{x},\Omega)\!=\!(0,0)$ and that of the upper is $(k_{x},\Omega)\!=\!(0,2\tfrac{|\alpha_{1}|}{\alpha_{2}})$. Increasing $\widetilde{v}$ further results in the upper branch approaching the lower branch until $\widetilde{v}\!=\!\widetilde{v}_{\mathrm{m}}$, whereupon $\alpha_{1}\!=\!0$ and $k_{x}\!=\!\pm\sqrt{\alpha_{2}}\Omega$; the spectrum becomes X-shaped as described above. Further increasing $\widetilde{v}\!>\!\widetilde{v}_{\mathrm{m}}$ produces once again a hyperbolic spectrum with a gap along $\Omega$, but the upper-branch vertex is $(k_{x},\Omega)\!=\!(0,0)$, whereas that of the lower is $(k_{x},\Omega)\!=\!(0,-2\tfrac{\alpha_{1}}{\alpha_{2}})$, which moves downwards with further increase in $\widetilde{v}$.

\subsection{Hyperbolic spectra}

The domain $\widetilde{v}\!>\!\widetilde{v}_{\mathrm{e}}$ ($\alpha_{2}\!>\!0$) where the spectrum is hyperbolic is divided into three distinct regions in terms of the sign of the determinant $\Delta\!=\!\alpha_{1}^{2}-\alpha_{0}\alpha_{2}$ of Eq.~\ref{Eq:kx_general} [Fig.~\ref{Fig:MTPlaneStructure}(e)].

The constraint $\Delta\!=\!0$ is satisfied along two curves that branch away from the point $(\widetilde{v},\varphi_{\mathrm{o}})\!=\!(\widetilde{v}_{\mathrm{m}},0^{\circ})$ in the MT-plane, and are identified as $\widetilde{v}_{\mathrm{min}}$ and $\widetilde{v}_{\mathrm{max}}$ in Fig.~\ref{Fig:MTPlaneStructure}(e), where:
\begin{equation}\label{Eq:VmaxVmin}
\widetilde{v}_{\mathrm{min}}=\widetilde{v}_{\mathrm{m}}\frac{\cos{\varphi_{\mathrm{B}}}}{\cos{(\varphi_{\mathrm{o}}-\varphi_{\mathrm{B}})}},\;\;\;\widetilde{v}_{\mathrm{max}}=\widetilde{v}_{\mathrm{m}}\frac{\cos{\varphi_{\mathrm{B}}}}{\cos{(\varphi_{\mathrm{o}}+\varphi_{\mathrm{B}})}}.
\end{equation}
Along either curve we have $k_{x}\!=\!\pm\sqrt{\alpha_{2}}(\Omega+\tfrac{\alpha_{1}}{\alpha_{2}})$, which is X-shaped with center at $\Omega\!=\!-\tfrac{\alpha_{1}}{\alpha_{2}}$. Changing $\varphi_{\mathrm{o}}$ along the $\widetilde{v}_{\mathrm{max}}$-curve results in $\widetilde{v}$ increasing monotonically, with the center of the X-spectrum moving \textit{downwards} along the $\Omega$-axis [Fig.~\ref{Fig:MTPlaneStructure}(f)]. Along the $\widetilde{v}_{\mathrm{min}}$-curve, on the other hand, $\widetilde{v}$ initially decreases with $\varphi_{\mathrm{o}}$ until $\varphi_{\mathrm{o}}\!=\!\varphi_{\mathrm{B}}$, after which $\widetilde{v}$ increases with $\varphi_{\mathrm{o}}$. In the first portion of the $\widetilde{v}_{\mathrm{min}}$-curve ($\varphi_{\mathrm{o}}\!<\!\varphi_{\mathrm{B}}$), the center of the X-spectrum moves \textit{upwards} along $\Omega$ [Fig.~\ref{Fig:MTPlaneStructure}(f)]. The center reaches infinity asymptotically at $\varphi_{\mathrm{o}}\!=\!\varphi_{\mathrm{B}}$, and the spectrum becomes separable with respect to $k_{x}$ and $\Omega$. In the second portion of the $\widetilde{v}_{\mathrm{min}}$-curve ($\varphi_{\mathrm{o}}\!>\!\varphi_{\mathrm{B}}$), the center of the X-spectrum starts at $\Omega\!\rightarrow\!-\infty$ and moves \textit{upwards} along $\Omega$ with increase in $\varphi_{\mathrm{o}}$. 

The region corresponding to a positive determinant $\Delta\!>\!0$ is formed of three disconnected areas in the MT-plane [Fig.~\ref{Fig:MTPlaneStructure}(e)]. These are the domains external to the two curves $\widetilde{v}_{\mathrm{min}}$ and $\widetilde{v}_{\mathrm{max}}$. In this case, we can write $k_{x}^{2}\!=\!\alpha_{2}(\Omega-\Omega_{+})(\Omega-\Omega_{-})$, and the spectrum is hyperbola with a gap along $\Omega$ whose vertices are $\Omega_{\pm}\!=\!\tfrac{-\alpha_{1}\pm\sqrt{\Delta}}{\alpha_{2}}$ [Fig.~\ref{Fig:MTPlaneStructure}(f)].

Finally, the regime where the determinant is negative $\Delta\!<\!0$ is the single connected domain enclosed by the branches $\widetilde{v}_{\mathrm{min}}$ and $\widetilde{v}_{\mathrm{max}}$ [Fig.~\ref{Fig:MTPlaneStructure}(e)]. There are no real zeros for $\alpha_{0}+2\alpha_{1}\Omega+\alpha_{2}\Omega^{2}\!=\!0$ in this domain, and the spectrum takes the form of a hyperbola with a gap along $k_{x}$ [Fig.~\ref{Fig:MTPlaneStructure}(f)].

\section{Experimental verification of spectral reorganization for space-time wave packets}

We proceed to verify the predictions outlined above concerning the spectral reorganization of STWPs in presence of normal GVD by synthesizing in free space STWPs that are propagation invariant once coupled to normally dispersive ZnSe. The ZnSe sample is in the form of stacked 1-inch-diameter discs each of thickness 5~mm (Thorlabs; WG71050) to a maximum thickness of 30~mm. The ZnSe Sellmeier equation is $n^{2}(\lambda)\!=\!4+\frac{1.9\lambda^{2}}{\lambda^{2}-0.113}$ ($\lambda$ in units of $\mu$m) \cite{Marple64JAP}. We select the operating wavelength $\lambda_{\mathrm{o}}\!\approx\!1$~$\mu$m, at which ZnSe is characterized by $\widetilde{v}_{\mathrm{m}}\!\approx\!0.39c$ (phase index $n_{\mathrm{m}}\!\approx\!2.49$ and group index $\widetilde{n}_{\mathrm{m}}\!\approx\!2.57$), and GVD coefficient $k_{2}\!\approx\!607$~fs$^2$/mm (corresponding to $\varphi_{\mathrm{B}}\!\approx\!19.2^{\circ}$).

We synthesize the STWPs starting from generic femtosecond pulses of width 100~fs at $\lambda_{\mathrm{o}}\!\sim\!1$~$\mu$m by associating with each wavelength $\lambda$ a prescribed spatial frequency $k_{x}(\lambda)$ according to Eq.~\ref{Eq:kx_general}, where $\lambda$ is the free-space wavelength. This is achieved by spatially resolving the pulse spectrum and then utilizing a spatial light modulator (SLM) to impart to each wavelength the appropriate propagation angle with respect to the $z$-axis. Because each wavelength is addressed independently, arbitrary forms of the relationship $k_{x}(\lambda)$ can be produced with such a strategy \cite{Hall21OEUniversal}. This setup therefore allows us to readily produce STWPs with any target values of $\widetilde{v}$ and $\varphi_{\mathrm{o}}$ within the paraxial regime by changing the 2D phase distribution imparted by the SLM to the spatially resolved wave front (see Appendix for details).

\begin{figure}[b!]
\centering
\includegraphics[width=7cm]{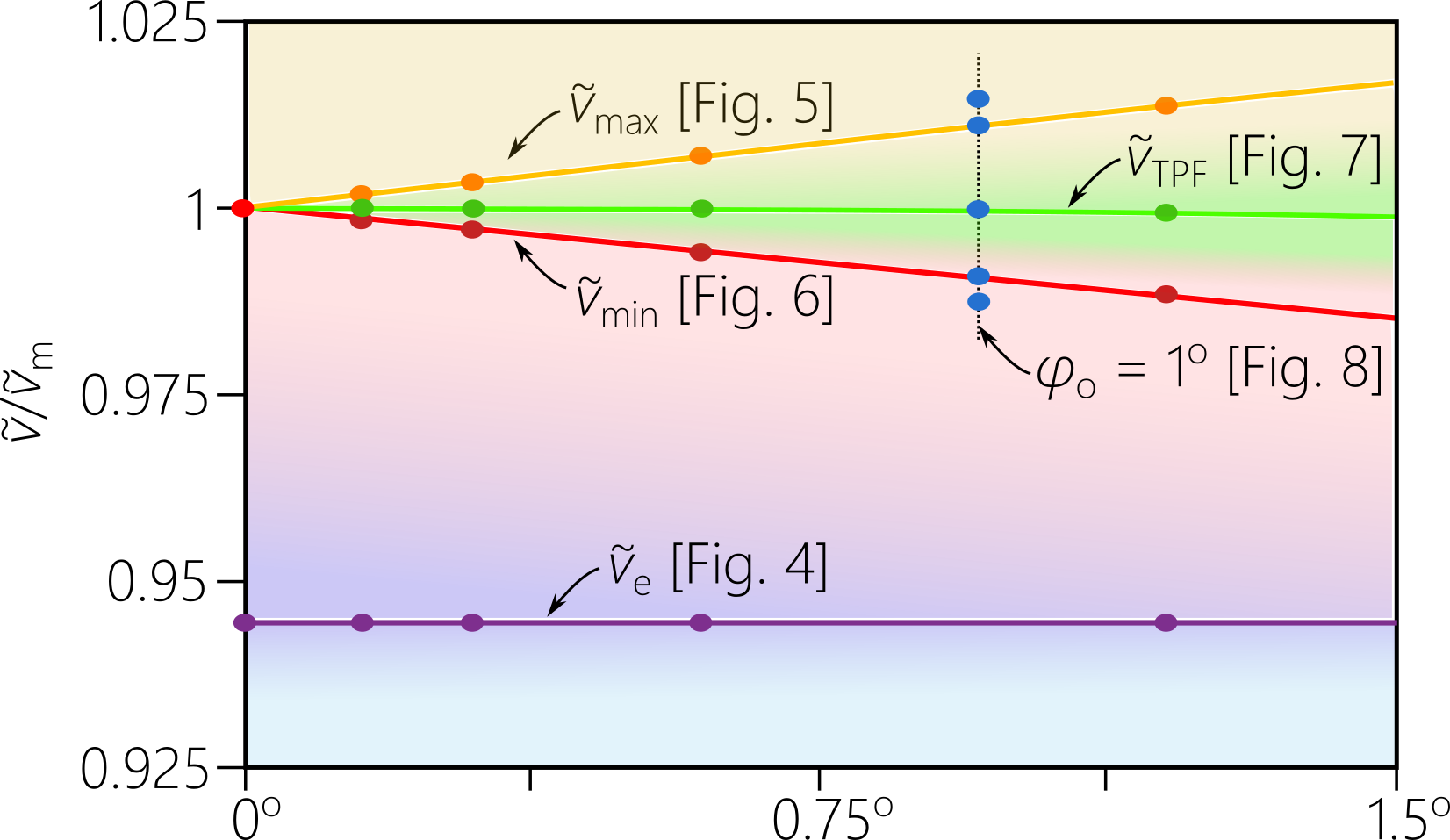}
\caption{Trajectories in the paraxial MT-plane ($\varphi_{\mathrm{o}}\!<\!1.5^{\circ}$) traced in our experiments. The points along the curves indicate the experimental values selected for $\widetilde{v}$ and $\varphi_{\mathrm{o}}$.}
\label{Fig:PointsUsed}
\end{figure}

At normal incidence onto the planar surface between free space and ZnSe, the transverse wave number $k_{x}$ is conserved because of translational invariance along $x$, and the optical frequency $\omega$ is invariant. Consequently the $(k_{x},\lambda)$-spectrum is a refractive invariant; however, the $(k_{z},\lambda)$-spectrum is \textit{not}. In the medium $k_{z}\!=\!\sqrt{(n(\omega)\tfrac{\omega}{c})^{2}-k_{x}^{2}}$ is linear in $\Omega$ in accordance with Eq.~\ref{Eq:GeneralConstraintKz} for all values of $\widetilde{v}$ and $\varphi_{\mathrm{o}}$, but in free space $k_{z}\!=\!\sqrt{(\tfrac{\omega}{c})^{2}-k_{x}^{2}}$ is not. In other words, the synthesized STWP is dispersion-free (to all orders) in ZnSe, but it undergoes GVD in free space. The change in the STWP group velocity upon transition from free space to the dispersive medium is governed by the law of refraction derived in \cite{He21Arxiv,Yessenov22OLDispersiveRefraction}.

The result is a STWP with profile $I(x,z;t)$ that is propagation invariant in ZnSe, $I(x,z;t)\!\rightarrow\!I(x,0;t-z/\widetilde{v})\!=\!I(x;\tau)$, where $\tau\!=\!t-z/\widetilde{v}$ is time in the moving frame of the STWP. For each STWP we measure the spectral projection onto the $(k_{x},\lambda)$-plane, from which we obtain the corresponding projection onto the $(k_{z},\lambda)$-plane. We also reconstruct the spatio-temporal intensity profile $I(x,z;\tau)\!=\!|\psi(x,z;\tau)|^{2}$ at fixed axial planes $z$ in the medium (see Appendix for details).

\begin{figure}[t!]
\centering
\includegraphics[width=7.3cm]{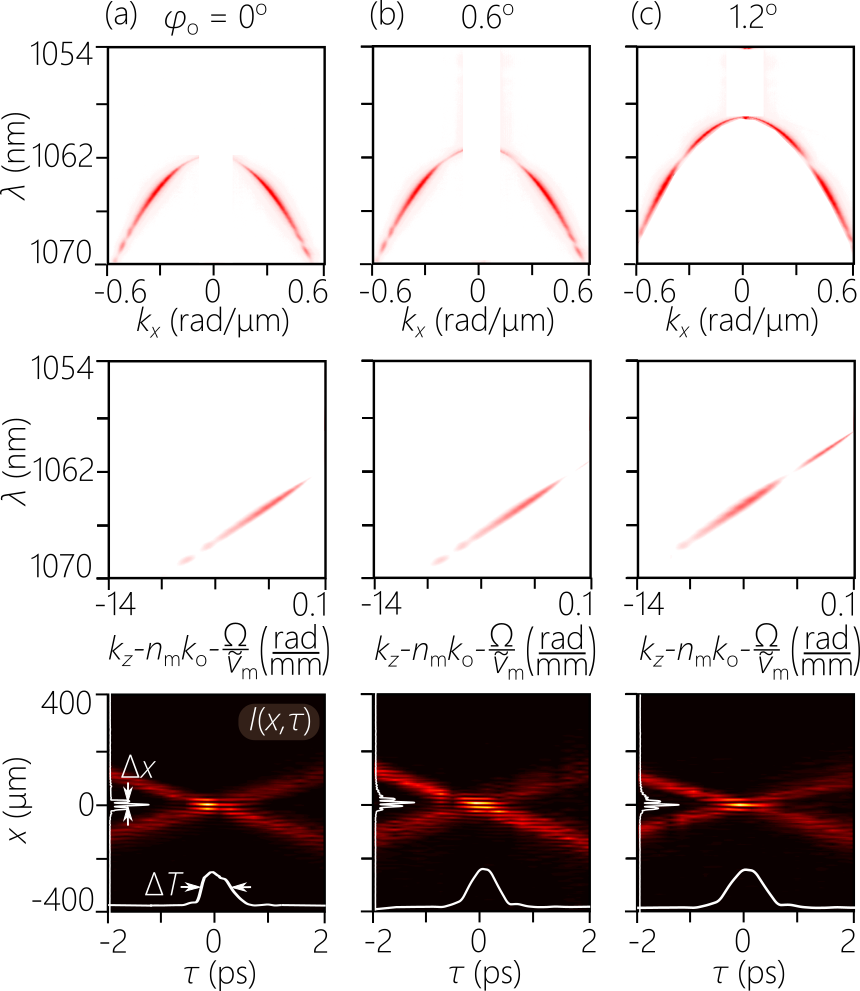}
\caption{Measured spectra and profiles for propagation invariant STWPs in presence of normal GVD while varying $\varphi_{\mathrm{o}}$ and holding $\widetilde{v}\!=\!\widetilde{v}_{\mathrm{e}}$. The first row is the spectrum projected onto the $(k_{x},\lambda)$-plane; the second row is the spectrum projected onto the $(k_{z},\lambda)$-plane in the medium; and the third row is the spatio-temporal intensity profile $I(x;\tau)$ measured at a fixed axial plane $z\!=\!30$~mm in ZnSe. Here $\tau\!=\!t-z/\widetilde{v}$ is time measured in the moving frame of the STWP. The white curve along the bottom horizontal axis is the on-axis pulse profile $I(0;\tau)$ of width $\Delta T$. The white curve along the left vertical axis is the beam profile at the pulse center $I(x;0)$ of width $\Delta x$. (a) $\varphi_{\mathrm{o}}\!=\!0^{\circ}$; (b) $\varphi_{\mathrm{o}}\!=\!0.6^{\circ}$; and (c) $\varphi_{\mathrm{o}}\!=\!1.2^{\circ}$.}
\label{Fig:escape}
\end{figure}

\begin{figure*}[t!]
\centering
\includegraphics[width=12cm]{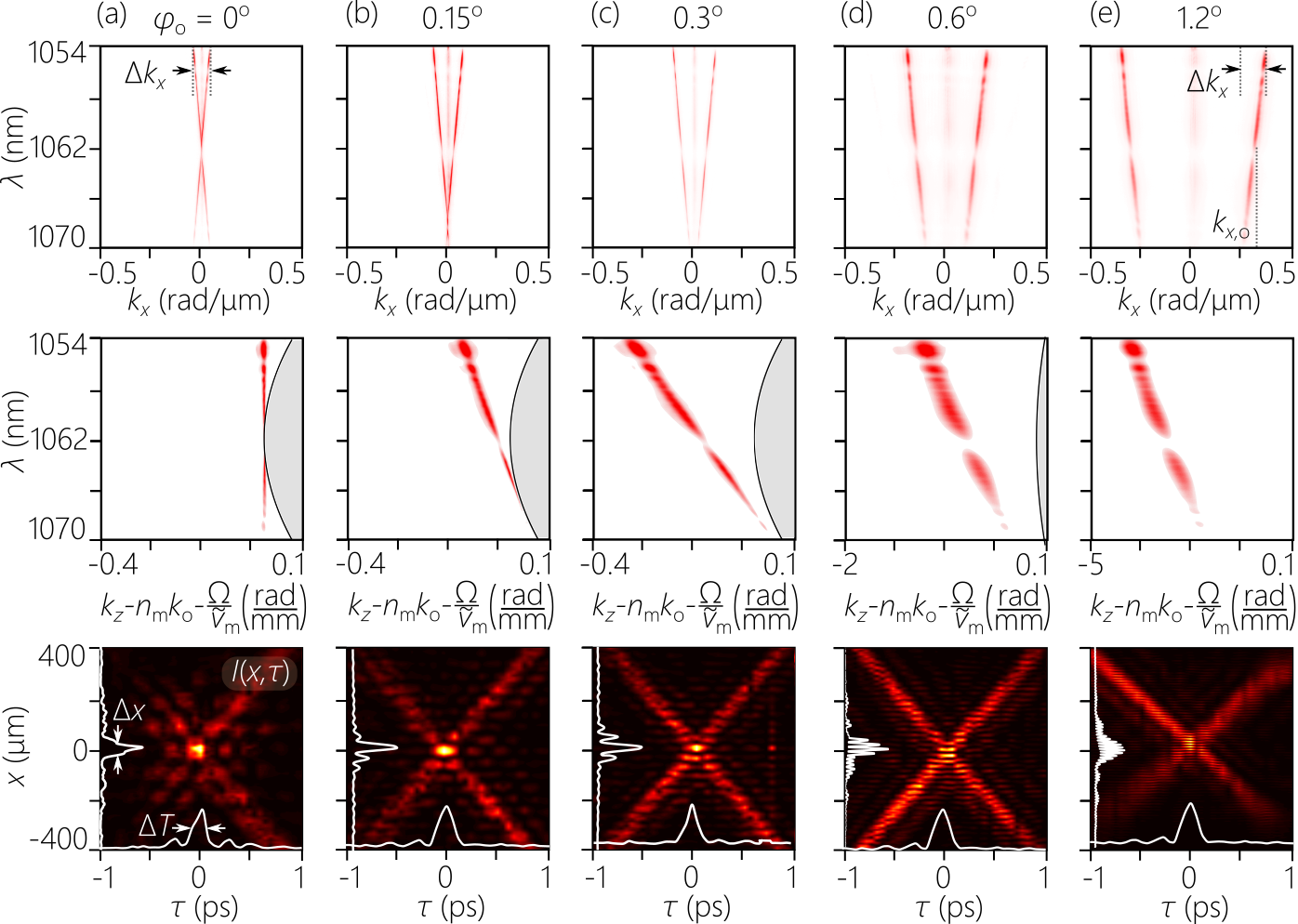}
\caption{Measured spectra and profiles for propagation-invariant STWPs in presence of normal GVD while varying $\varphi_{\mathrm{o}}$ along the curve $\widetilde{v}_{\mathrm{max}}$ [Fig.~\ref{Fig:MTPlaneStructure}(e)]. The rows have the same structure as in Fig.~\ref{Fig:escape}. The curves in the second row panels are the dispersive light-line in the medium ($k_{x}\!=\!0$), and the shaded area corresponds to evanescent waves. (a) $\varphi_{\mathrm{o}}\!=\!0^{\circ}$; (b) $\varphi_{\mathrm{o}}\!=\!0.15^{\circ}$; (c) $\varphi_{\mathrm{o}}\!=\!0.3^{\circ}$; (d) $\varphi_{\mathrm{o}}\!=\!0.6^{\circ}$; and (e) $\varphi_{\mathrm{o}}\!=\!1.2^{\circ}$.}
\label{Fig:Vmax}
\end{figure*}

\begin{figure*}[t!]
\centering
\includegraphics[width=12cm]{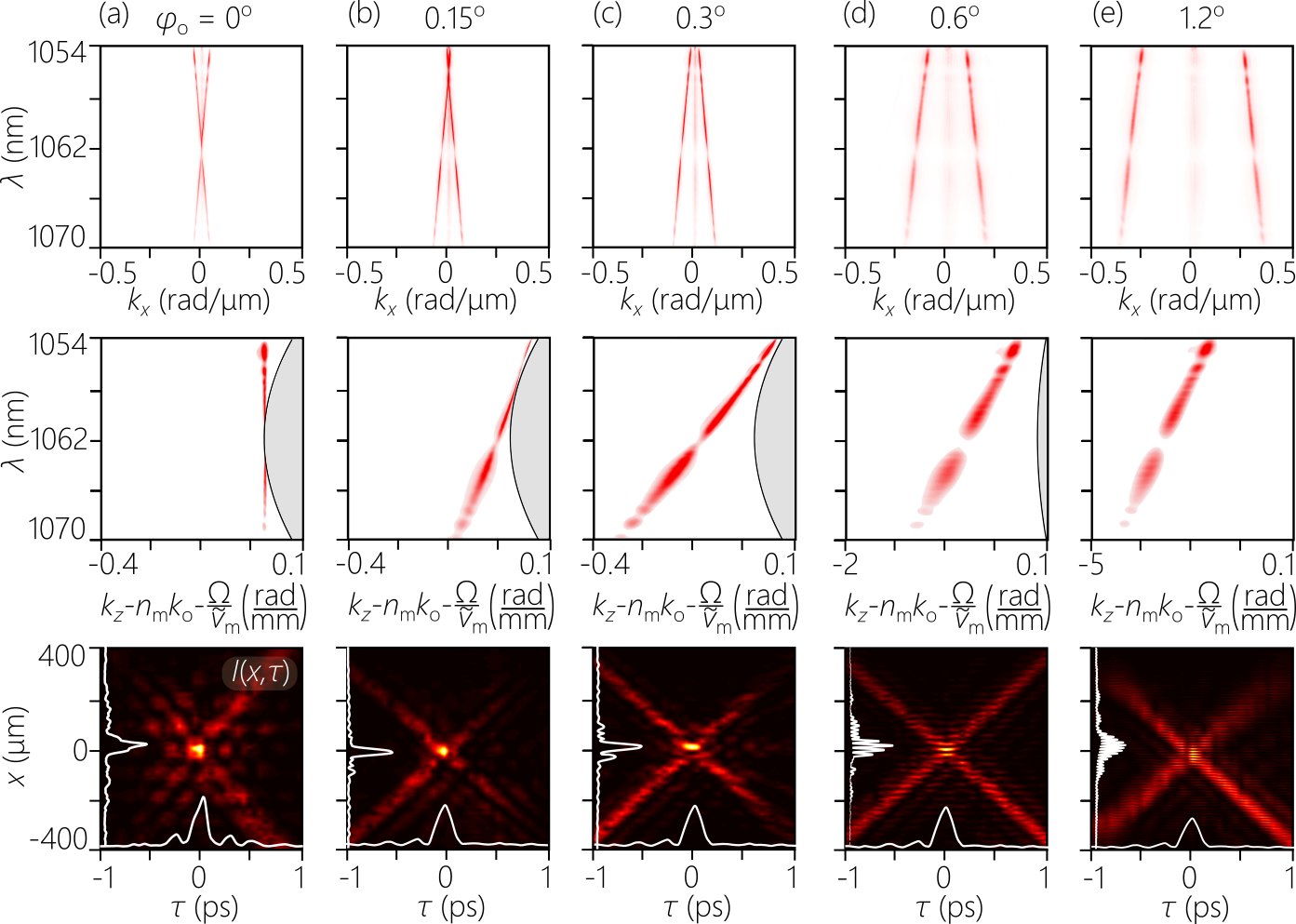}
\caption{Measured spectra and profiles for propagation-invariant STWPs in presence of normal GVD while varying $\varphi_{\mathrm{o}}$ along the curve $\widetilde{v}_{\mathrm{min}}$ [Fig.~\ref{Fig:MTPlaneStructure}(e)]. The rows have the same structure as in Fig.~\ref{Fig:escape}. (a) $\varphi_{\mathrm{o}}\!=\!0^{\circ}$; (b) $\varphi_{\mathrm{o}}\!=\!0.15^{\circ}$; (c) $\varphi_{\mathrm{o}}\!=\!0.3^{\circ}$; (d) $\varphi_{\mathrm{o}}\!=\!0.6^{\circ}$; and (e) $\varphi_{\mathrm{o}}\!=\!1.2^{\circ}$.}
\label{Fig:Vmin}
\end{figure*}

The synthesized STWPs have a bandwidth in the range $\Delta\lambda\!=\!8-16$~nm at a wavelength $\lambda_{\mathrm{o}}\!\approx\!1062$~nm, corresponding to pulsewidths of $\Delta T\!\approx\!200-450$~fs for the on-axis profile $I(0;\tau)$; we denote the pulsewidth at the beam center $\Delta T$. The beam width $\Delta x$ at the pulse center $I(x;0)$ is determined by the spatial bandwidth $\Delta k_{x}$, which is related to the temporal bandwidth $\Delta\lambda$ through the STWP parameters $\widetilde{v}$ and $\varphi_{\mathrm{o}}$ (via Eq.~\ref{Eq:kx_general}). In the measured spectra, we plot the results with $\lambda$ decreasing upwards along the vertical axis to retain the orientation of the curves associated with $\Omega$ as shown in Fig.~\ref{Fig:MTPlaneStructure}. Furthermore, we plot the spectra projected onto the $(k_{z},\lambda)$-plane with $k_{z}-n_{\mathrm{m}}k_{\mathrm{o}}-\Omega/\widetilde{v}_{\mathrm{m}}$ along the horizontal axis for clarity. Although these spectra are linear only in the $(k_{z},\Omega)$-plane, they remain approximately linear in the $(k_{z},\lambda)$-plane because $\Delta\lambda\!\ll\!\lambda_{\mathrm{o}}$.  We plot the profiles at the maximum ZnSe sample length of 30~mm, but the profile remains invariant at shorter ZnSe lengths [Fig.~\ref{Fig:PropagationInvariance}].

\begin{figure*}[t!]
\centering
\includegraphics[width=12cm]{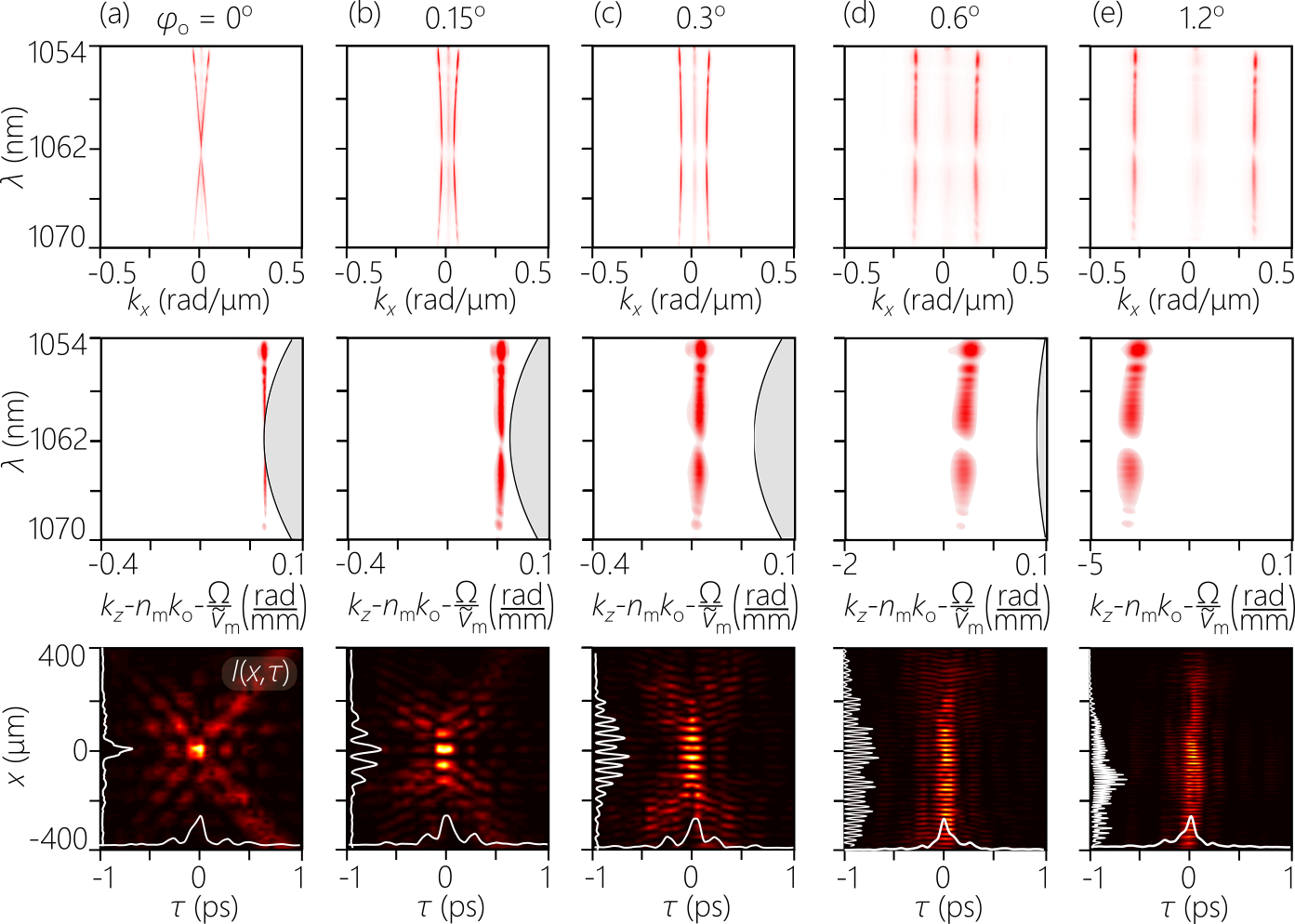}
\caption{Measured spectra and profiles for propagation-invariant STWPs in presence of normal GVD while varying $\varphi_{\mathrm{o}}$ and simultaneously maintaining $\widetilde{v}\!=\!\widetilde{v}_{\mathrm{TPF}}(\varphi_{\mathrm{o}})$. The rows have the same structure as in Fig.~\ref{Fig:escape}. (a) $\varphi_{\mathrm{o}}\!=\!0^{\circ}$; (b) $\varphi_{\mathrm{o}}\!=\!0.15^{\circ}$; (c) $\varphi_{\mathrm{o}}\!=\!0.3^{\circ}$; (d) $\varphi_{\mathrm{o}}\!=\!0.6^{\circ}$; and (e) $\varphi_{\mathrm{o}}\!=\!1.2^{\circ}$.}
\label{Fig:vs}
\end{figure*}

We trace along several of the crucial trajectories in the MT-plane that bring out the rich dynamics of spectral reorganization: varying $\varphi_{\mathrm{o}}$ while $\widetilde{v}\!=\!\widetilde{v}_{\mathrm{e}}$ [Fig.~\ref{Fig:escape}], $\widetilde{v}\!=\!\widetilde{v}_{\mathrm{max}}$ [Fig.~\ref{Fig:Vmax}]; $\widetilde{v}\!=\!\widetilde{v}_{\mathrm{min}}$ [Fig.~\ref{Fig:Vmin}], $\widetilde{v}\!=\!\widetilde{v}_{\mathrm{TPF}}(\varphi_{\mathrm{o}})$ [Fig.~\ref{Fig:vs}], and varying $\widetilde{v}$ while holding $\varphi_{\mathrm{o}}\!=\!1^{\circ}$ [Fig.~\ref{Fig:fixedphi}]. We identify in Fig.~\ref{Fig:PointsUsed} the points along those trajectories. In our experiments we remain within the paraxial limit with $\varphi_{\mathrm{o}}\!<\!2^{\circ}$, which in turn requires that $\widetilde{v}$ remains close to $\widetilde{v}_{\mathrm{m}}$. Throughout, the measured STWP spectra and profiles are in excellent agreement with calculations based on the theoretical model described in the previous Section without any fitting parameters.

\subsection{Escape velocity}

In ZnSe at $\lambda_{\mathrm{o}}\!\approx\!1$~$\mu$m, the escape velocity is $\widetilde{v}_{\mathrm{e}}\!\approx\!0.944\widetilde{v}_{\mathrm{m}}\!\approx\!0.367c$. In our first experiment [Fig.~\ref{Fig:escape}], we hold $\widetilde{v}$ fixed at $\widetilde{v}\!=\!\widetilde{v}_{\mathrm{e}}$ ($\alpha_{2}\!=\!0$) and vary $\varphi_{\mathrm{o}}$ from $0^{\circ}$ to $1.2^{\circ}$. The spectrum is a parabola that is symmetric around $k_{x}\!=\!0$ and whose vertex along the $\Omega$-axis moves from $\Omega\!=\!0$ upwards as $\varphi_{\mathrm{o}}$ increases. The STWP profile remains X-shaped throughout. Because the temporal bandwidth $\Delta\lambda$ and the spatial bandwidth $\Delta k_{x}$ both remain almost constant while $\varphi_{\mathrm{o}}$ is varied, the on-axis pulsewidth $\Delta T$ and the beam width $\Delta x$ at the pulse center in turn remain almost constant.  

\subsection{Hyperbolic spectra with positive and negative determinants}

Next, we vary $\varphi_{\mathrm{o}}$ and simultaneously tune $\widetilde{v}$ to produce STWPs represented by the points along the curves corresponding to $\widetilde{v}_{\mathrm{max}}$ and $\widetilde{v}_{\mathrm{min}}$ (Eq.~\ref{Eq:VmaxVmin}) identified in Fig.~\ref{Fig:PointsUsed}. The spectra for the STWPs associated with $\widetilde{v}_{\mathrm{max}}$ [Fig.~\ref{Fig:Vmax}] are quite similar to those associated with $\widetilde{v}_{\mathrm{min}}$ [Fig.~\ref{Fig:Vmin}] except that they are inverted along the $\lambda$-axis. At $\varphi_{\mathrm{o}}\!=\!0^{\circ}$ where $\widetilde{v}\!=\!\widetilde{v}_{\mathrm{m}}$ (a point common to both $\widetilde{v}_{\mathrm{max}}$ and $\widetilde{v}_{\mathrm{min}}$), the spectrum is X-shaped with center at $(k_{x},\Omega)\!=\!(0,0)$. Increasing $\varphi_{\mathrm{o}}$ retains the overall structure except that the intersection point moves away from $(k_{x},\Omega)\!=\!(0,0)$, and at $\varphi_{\mathrm{o}}\!>\!0.3^{\circ}$, the intersection point exits the accessible spectral range. The temporal bandwidth remains fixed throughout ($\Delta\lambda\!\approx\!16$~nm), so that the on-axis pulsewidth $\Delta T$ also remains fixed. However, the reorganization in the spatio-temporal spectral structure leads to a noticeable change in the spatial beam profile. First, as $\varphi_{\mathrm{o}}$ increases [Fig.~\ref{Fig:Vmax}(b,c) and Fig.~\ref{Fig:Vmin}(b,c)], the spatial bandwidth $\Delta k_{x}$ increases, and thus the beam width $\Delta x$ becomes narrower. However, with $\varphi_{\mathrm{o}}\!>\!0.3^{\circ}$ [Fig.~\ref{Fig:Vmax}(d,e) and Fig.~\ref{Fig:Vmin}(d,e)], the spatial spectrum is no longer `baseband', and is instead centered at a non-zero-valued spatial frequency $k_{x,\mathrm{o}}$. The region in the vicinity of $k_{x}\!=\!0$ is eliminated from the spatial spectrum, so that the envelope of the beam profile at $\tau\!=\!0$ is accompanied by spatial fringes whose period decreases with $\varphi_{\mathrm{o}}$. Furthermore, broadening of the spatial envelope width $\Delta x$ occurs because the spatial bandwidth $\Delta k_{x}$ decreases with $\varphi_{\mathrm{o}}$ [Fig.~\ref{Fig:Vmax}(d,e) and Fig.~\ref{Fig:Vmin}(d,e)]. 

\subsection{TPF group velocity}

Next, we examine the STWPs whose spectra are represented by points along the curve $\widetilde{v}\!=\!\widetilde{v}_{\mathrm{TPF}}(\varphi_{\mathrm{o}})$, where $\alpha_{1}\!=\!0$, over the span from $\varphi_{\mathrm{o}}\!=\!0^{\circ}$ to $1.2^{\circ}$. At $\varphi_{\mathrm{o}}\!=\!0^{\circ}$, $\alpha_{0}\!=\!\alpha_{1}\!=\!0$, $\alpha_{2}\!>\!0$, and $k_{x}\!=\!\pm\sqrt{\alpha_{2}}\Omega$, which is an X-shaped spectrum and the profile $I(x;\tau)$ is clearly X-shaped [Fig.~\ref{Fig:vs}(a)]. Increasing $\varphi_{\mathrm{o}}$ results in the X-shaped spectrum morphing into a hyperbola with a gap along the $k_{x}$-axis. The profile $I(x;\tau)$ is X-shaped, but whose branches have started to lose their distinctiveness, and spatial oscillations appear [Fig.~\ref{Fig:vs}(b)]. The curvature of the hyperbola drops with increasing $\varphi_{\mathrm{o}}$, so that $k_{x}$ and $\Omega$ decouple, and the profile $I(x;\tau)$ is no longer X-shaped [Fig.~\ref{Fig:vs}(c-e)]. Rather, it becomes approximately separable along space and time, $\psi(x,0;t)\!\approx\!\psi_{x}(x)\psi_{t}(t)$, with $\psi_{x}(x)$ approaching a cosine function $\psi_{x}(x)\!\approx\!\cos{k_{x}x}$, and $\psi_{t}(t)$ a transform-limited temporal pulse corresponding to the bandwidth $\Delta\lambda$. This spatio-temporal separability becomes exact when $\varphi_{\mathrm{o}}\!=\!\varphi_{\mathrm{B}}$, which occurs in the non-paraxial regime. The measurements in Fig.~\ref{Fig:vs}(d,e) thus approximate in the transverse dimension the predicted propagation invariance of pulsed Bessel beams \cite{Liu98JMO,Hu02JOSAA,Lu03JOSAA,Turunen10PO}. Note that the on-axis pulse width $\Delta T$ is constant throughout because $\Delta\lambda$ is maintained fixed.

\begin{figure*}[t!]
\centering
\includegraphics[width=12cm]{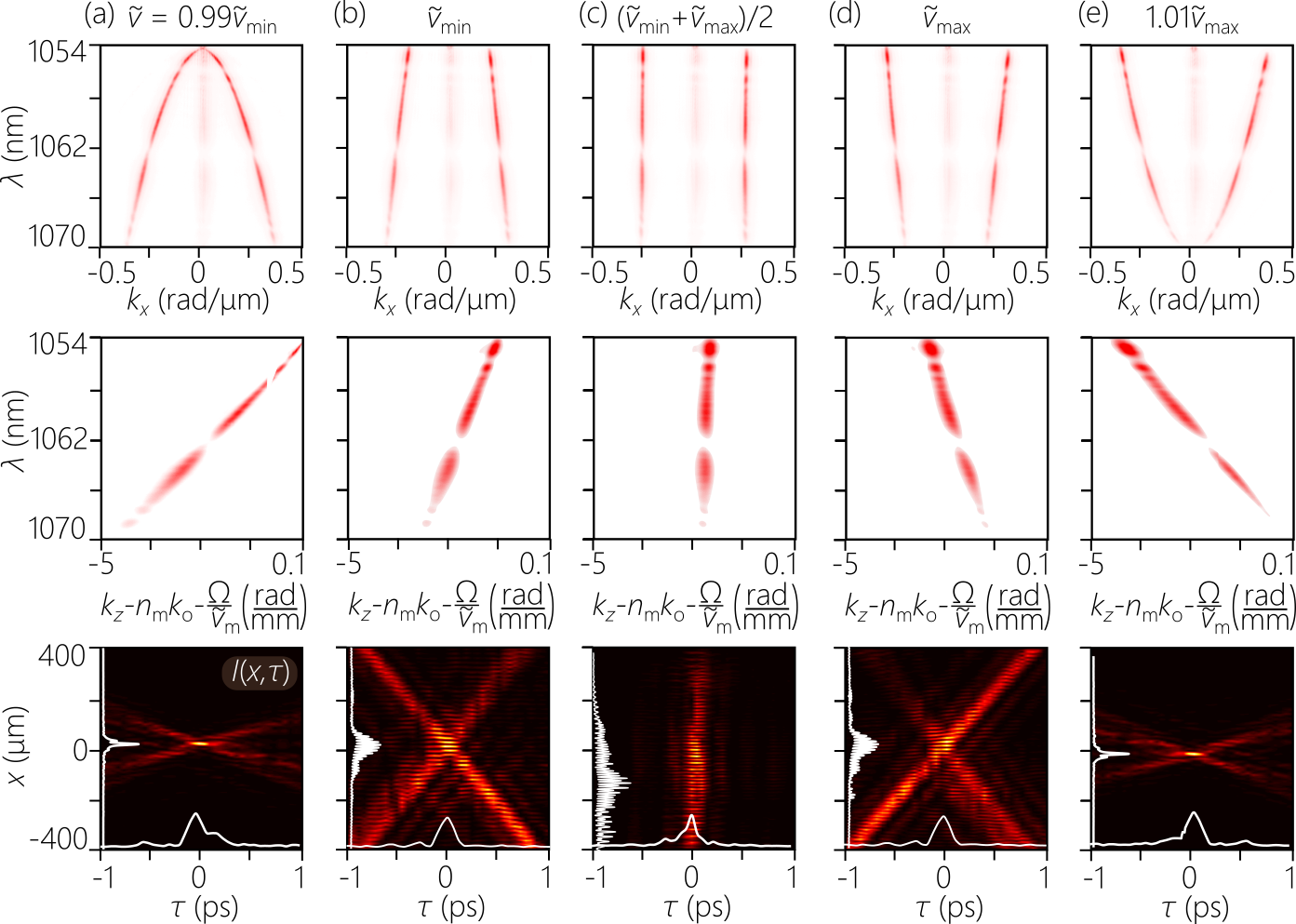}
\caption{Measured spectra and profiles for propagation-invariant STWPs in presence of normal GVD while varying $\widetilde{v}$ and holding $\varphi_{\mathrm{o}}\!=\!1^{\circ}$. The rows have the same structure as in Fig.~\ref{Fig:escape}. (a) $\widetilde{v}\!=\!0.99\widetilde{v}_{\mathrm{min}}$; (b) $\widetilde{v}\!=\!\widetilde{v}_{\mathrm{min}}$; (c) $\widetilde{v}\!=\!(\widetilde{v}_{\mathrm{min}}+\widetilde{v}_{\mathrm{max}})/2$; (d) $\widetilde{v}\!=\!\widetilde{v}_{\mathrm{max}}$; and (e) $\widetilde{v}\!=\!1.01\widetilde{v}_{\mathrm{max}}$. Here $\widetilde{v}_{\mathrm{min}}\!=\!0.99\widetilde{v}_{\mathrm{m}}$, $\widetilde{v}_{\mathrm{max}}\!=\!1.01\widetilde{v}_{\mathrm{m}}$, and $\widetilde{v}_{\mathrm{m}}\!\approx\!0.39c$.}
\label{Fig:fixedphi}
\end{figure*}

\subsection{Changing the group velocity at fixed angular offset}

Finally, we examine the spectral reorganization as we tune $\widetilde{v}$ while holding $\varphi_{\mathrm{o}}$ fixed. At $\varphi_{\mathrm{o}}\!=\!1^{\circ}$ [Fig.~\ref{Fig:PointsUsed}], realizing the group velocity $\widetilde{v}\!=\!0.99\widetilde{v}_{\mathrm{min}}\!>\!\widetilde{v}_{\mathrm{e}}$ results in a hyperbolic spectrum with gap along $\Omega$, although only one branch of it fits in the selected spectral window. Increasing $\widetilde{v}$ to $\widetilde{v}_{\mathrm{min}}$ results in an X-spectrum whose center is outside the spectral window. Whereas $\Delta\lambda$ remains fixed, this spectral reorganization is accompanied by a decrease in $\Delta k_{x}$ and a displacement of the spatial-spectrum center to a non-zero spatial frequency. Therefore, the beam width $\Delta x$ increases, and is accompanied by rapid spatial oscillations [Fig.~\ref{Fig:fixedphi}(b)]. Furthermore, the tilt angle of the two branches of the X-shaped profile (the so-called pulse-front tilt \cite{Hebling96OQE}) changes dramatically as a result of the change in curvature of the spatio-temporal spectrum \cite{Hall21OL}. For $\widetilde{v}_{\mathrm{min}}\!<\!\widetilde{v}\!<\!\widetilde{v}_{\mathrm{max}}$, the spectrum is a hyperbola with gap along $k_{x}$. The hyperbolic curvature here is small, so that the spectrum is approximately separable along $k_{x}$ and $\Omega$, and the profile is also approximately separable with respect to space and time (a cosine wave along $x$ and a pulse whose width is determined by $\Delta\lambda$ along $t$); see Fig.~\ref{Fig:fixedphi}(c). Increasing $\widetilde{v}$ further reverses these changes and recapitulated the earlier spectral reorganization but with the spectra flipped along the $\lambda$-axis [Fig.~\ref{Fig:fixedphi}(d,e)].

\section{Discussion}

The STWPs that we have studied are all undergirded by angular dispersion \cite{Torres10AOP,Fulop10Review}; that is, each frequency $\omega$ travels in the medium at a different angle $\varphi(\omega)$ with respect to the $z$-axis, with $k_{z}(\omega)\!=\!\tfrac{\omega}{c}\cos{\varphi(\omega)}$ and $k_{x}(\omega)\!=\!\tfrac{\omega}{c}\sin{\varphi(\omega)}$. To achieve dispersion-free propagation in the normal-GVD regime, the angular dispersion introduced into the field must yield anomalous GVD that counterbalances the normal GVD in the medium. It is well-established that angular dispersion does indeed introduce anomalous GVD as needed here \cite{Martinez84JOSAA,Fork84OL,Gordon84OL}. However, as we recently uncovered \cite{Hall21OL,Hall21OL3NormalGVD,Hall22OEConsequences}, the proof provided by Martinez, Gordon, and Fork \cite{Martinez84JOSAA} tacitly assumes that the angular dispersion is differentiable; that is, $\tfrac{d\varphi}{d\omega}$ is defined everywhere. These field configurations are typically called TPFs, and correspond to the curve $\widetilde{v}\!=\!\widetilde{v}_{\mathrm{TPF}}(\varphi_{\mathrm{o}})$ in Fig.~\ref{Fig:MTPlaneStructure}(c). We have shown that angular dispersion can only yield normal GVD (thus allowing propagation invariance in the anomalous GVD regime) when $\tfrac{d\varphi}{d\omega}$ is \textit{not} defined at some frequency (\textit{non-differentiable} angular dispersion) \cite{Hall21OL3NormalGVD,Hall22OEConsequences}. However, non-differentiable angular dispersion can \textit{also} produce dispersion-free STWPs in the normal-GVD regime. Many of the STWPs examined here correspond to non-differentiable angular dispersion; e.g., all the hyperbolic spectra with a gap along $\Omega$ in which $\tfrac{d\varphi}{d\Omega}$ is not be defined at the vertices of the two branches of the hyperbola. More theoretical work is needed to fully identify the regions of differentiable and non-differentiable angular dispersion in the MT-plane in presence of normal GVD.

We have identified one point in the MT-plane at $(\widetilde{v},\varphi_{\mathrm{o}})\!=\!(\widetilde{v}_{\mathrm{e}},\varphi_{\mathrm{B}})$ that corresponds to a pulsed Bessel beam $I_{\mathrm{B}}(x;\tau)$ that separates into a product of spatial and temporal profiles $I_{\mathrm{B}}\!\propto\!J_{0}^{2}(k_{r}r)|\psi_{t}(t-z/\widetilde{v}_{\mathrm{e}})|^{2}$. Although the spectrum is separable with respect to $k_{x}$ and $\Omega$, the field nevertheless incorporates angular dispersion: $k_{x}\!=\!\pm n(\omega)\tfrac{\omega}{c}\sin{\varphi(\omega)}$. Each frequency $\omega$ travels at an angle $\varphi(\omega)$ with respect to the $z$-axis in the medium in such a way that $k_{x}(\Omega)$ is held constant. Another propagation-invariant separable wave packet is the Bessel-Airy wave packet \cite{Chong10NP} in which the pulse profile is an Airy function, which accelerates in a dispersive medium (whether normal or anomalous GVD). In contrast, the pulsed Bessel beam can have an arbitrary temporal waveform; however, this applies only in the normal-GVD regime, and there are no anomalous-GVD counterparts.

\begin{figure}[t!]
\centering
\includegraphics[width=7.3cm]{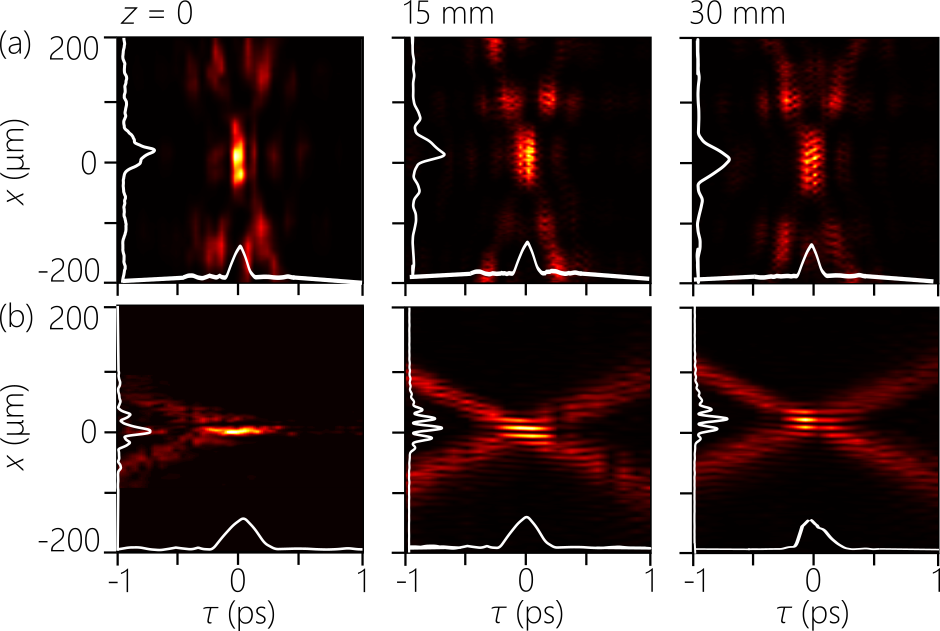}
\caption{Verifying the propagation invariance of the STWPs in presence of normal GVD over the course of 30-mm propagation distance in ZnSe. (a) The STWP corresponding to the point $(\widetilde{v},\varphi_{\mathrm{o}})\!=\!(\widetilde{v}_{\mathrm{m}},0^{\circ})$ and (b) $(\widetilde{v},\varphi_{\mathrm{o}})\!=\!(\widetilde{v}_{\mathrm{e}},0^{\circ})$ in the MT-plane.}
\label{Fig:PropagationInvariance}
\end{figure}

Finally, we emphasize that all the spatio-temporal profiles depicted in Fig.~\ref{Fig:escape} through Fig.~\ref{Fig:fixedphi} are propagation invariant -- they are all independent of $z$ in the medium; see Fig.~\ref{Fig:PropagationInvariance}.

\section{Conclusion}

In conclusion, we have explored the rich dynamics of spectral reorganization for propagation-invariant STWPs in presence of normal GVD. Two parameters identify each family of STWPs, the group velocity and the central axial wave number. By varying these two parameters, a wide range of spatio-temporal spectra are produced that are compatible with propagation invariance in the dispersive medium, including X-shaped, hyperbolic (with gaps along the spatial-frequency or temporal-frequency axes), parabolic, or elliptic, or separable spatio-temporal spectra. All these spectra result from the intersection of the dispersive light-cone with a spectral plane that is tilted and displaced with respect to the light-cone. We have experimentally validated these dynamics of spectral reorganization for STWPs in the near-infrared at a wavelength $\sim\!1$~$\mu$m with pulses of width $\sim\!200-450$~fs and bandwidth $\Delta\lambda\!\approx\!8-16$~nm in the paraxial regime using ZnSe in its normal-GVD regime. These results pave the way to exploiting STWPs in initiating nonlinear optical processes by accessing new forms of phase-matching that are engendered by their unique spatio-temporal spectral structures. 

\begin{figure}[t!]
\centering
\includegraphics[width=8.6cm]{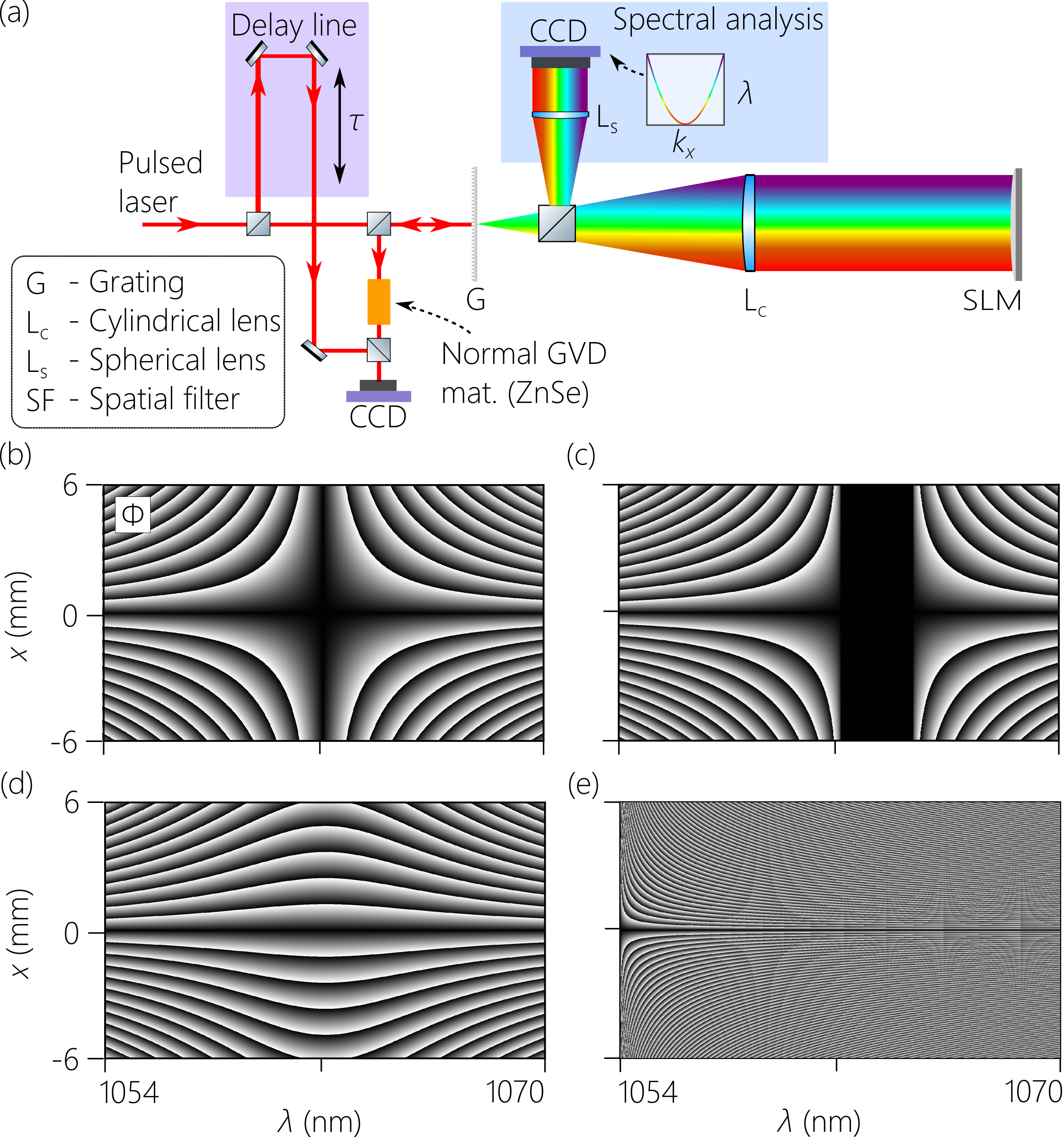}
\caption{(a) Schematic of the universal angular-dispersion synthesizer used to prepare the STWPs reported on here. (b-e) Examples of 2D phase patterns $\Phi(\lambda,x)$ imparted by the SLM to the incident spectrally resolved wave front. The phase patterns produce STWPs corresponding to the following points in the MT-plane: (b) $(\widetilde{v},\varphi_{\mathrm{o}})\!=\!(\widetilde{v}_{\mathrm{m}},0^{\circ})$, (c) $(\widetilde{v},\varphi_{\mathrm{o}})\!=\!(1.001\widetilde{v}_{\mathrm{m}},0^{\circ})$, (d) $(\widetilde{v},\varphi_{\mathrm{o}})\!=\!(\widetilde{v}_{\mathrm{TPF}},0.2^{\circ})$, and (e) $(\widetilde{v},\varphi_{\mathrm{o}})\!=\!(\widetilde{v}_{\mathrm{e}},0^{\circ})$.}
\label{Fig:Setup}
\end{figure}

\section*{Appendix: Experimental details}

The setup used to synthesize and characterize the STWPs utilizes the universal angular-dispersion synthesizer described in \cite{Hall21OEUniversal} as shown in Fig.~\ref{Fig:Setup}(a). Plane-wave pulses (width $\approx\!100$~fs, bandwidth $\approx\!25$~nm, at a central wavelength of $\lambda_{\mathrm{o}}\!\approx\!1064$~nm) are spectrally resolved via a diffraction grating (1200~lines/mm), collimated by a cylindrical lens (focal length $f\!=\!500$~mm), and modulated by a phase-only spatial light modulator (SLM; Meadowlark, E19X12) placed at the focal plane of the lens. Each wavelength is confined to a column on the SLM where it acquires the phase $\Phi(x,\lambda)\!=\!\pm\frac{2\pi}{\lambda}\sin{\{\varphi'(\lambda)\}}x$, where $\varphi'(\lambda)$ is a deflection angle for $\lambda$ with respect to the $z$-axis in free space (and is related to the angle $\varphi(\lambda)$ in the medium via Snell's law). The retro-reflected wave front returns to the grating whereupon the STWP is reconstituted. Examples of the 2D phase distributions imparted to the spectrally resolved wave fronts are shown in Fig.~\ref{Fig:Setup}(b-e). The phase distribution in Fig.~\ref{Fig:Setup}(b) corresponds to the X-shaped spectrum associated with STWP represented by the point $(\widetilde{v},\varphi_{\mathrm{o}})\!=\!(\widetilde{v}_{\mathrm{m}},0^{\circ})$ in the MT-plane. Moving to the point $(\widetilde{v},\varphi_{\mathrm{o}})\!=\!(1.001\widetilde{v}_{\mathrm{m}},0^{\circ})$ in Fig.~\ref{Fig:Setup}(c), the spectrum is hyperbolic with a gap along the $\Omega$-axis, which is manifest in the portion of the phase distribution that is nulled. In contrast, the phase distribution in Fig.~\ref{Fig:Setup}(d) is associated with STWP represented by the point $(\widetilde{v},\varphi_{\mathrm{o}})\!=\!(\widetilde{v}_{\mathrm{TPF}},0.2^{\circ})$, which corresponds to a hyperbolic spectrum but with the gap along $k_{x}$ rather than $\Omega$. Finally, the phase distribution Fig.~\ref{Fig:Setup}(e) is associated with STWP represented by the point $(\widetilde{v},\varphi_{\mathrm{o}})\!=\!(\widetilde{v}_{\mathrm{e}},0^{\circ})$, which corresponds to a parabolic spectrum at the escape velocity. Here $k_{x}$ changes monotonically with $\lambda$.

The spatio-temporal spectrum in the $(k_{x},\lambda)$-plane is acquired by introducing a spatial Fourier transform in the path of the spectrally resolved wave front and recording the intensity with a CCD camera \cite{Kondakci17NP}. From the spectrum in the $(k_{x},\lambda)$-plane, we obtain the associated spectrum in the $(k_{z},\lambda)$-plane for the field in the medium from $k_{z}^{2}\!=\!(n(\omega)\frac{\omega}{c})^{2}-k_{x}^{2}$, which is to be compared with the theoretical expectation in Eq.~\ref{Eq:GeneralConstraintKz}. The spatio-temporal intensity profile of the STWP $I(x,z;\tau)$ is reconstructed at a given axial plane $z$ by interfering it with the initial laser pulses as a reference \cite{Kondakci19NC,Yessenov19OE,Bhaduri20NP}; see Fig.~\ref{Fig:Setup}(a). When the reference pulse and the STWP overlap in space and time, the visibility of spatially resolved fringes recorded by a CCD camera helps reconstruct the STWP profile as we sweep an optical delay $\tau$ placed in the reference path. 

We verify in Fig.~\ref{Fig:PropagationInvariance} the propagation invariance of the STWPs in ZnSe. We reconstruct the intensity profile $I(x,z;\tau)$ at three axial planes: $z\!=\!0$ before entrance to the ZnSe sample; $z\!=\!15$~mm; and $z\!=\!30$~mm. We plot the results for two different STWPs corresponding to the points $(\widetilde{v},\varphi_{\mathrm{o}})\!=\!(\widetilde{v}_{\mathrm{m}},0^{\circ})$ and $(\widetilde{v},\varphi_{\mathrm{o}})\!=\!(\widetilde{v}_{\mathrm{e}},0^{\circ})$ in the MT-plane. In both cases, the profile remains unchanged with propagation in the dispersive medium.

\section*{ACKNOWLEDGMENTS}

The authors acknowledge the support of the Office of Naval Research (ONR, Grant Nos. N00014-17-1-2458 and N00014-20-1-2789).

The authors declare no conflicts of interest.

\section*{DATA AVAILABILITY}

The data that support the findings of this study are available from the corresponding author upon reasonable request.

\bibliography{diffraction}

\end{document}